\documentclass[aps,prx,twocolumn,superscriptaddress,showpacs,final]{revtex4-2} 
\usepackage{amssymb}
\usepackage{suffix}
\usepackage{mathtools}
\usepackage[utf8]{inputenc}
\usepackage{booktabs}
\usepackage{cases}
\usepackage[multiple]{footmisc}
\usepackage{dcolumn}
\usepackage{color,soul}
\usepackage{rotating}
\usepackage{perpage}
\usepackage{xcolor}
\usepackage{soul}
\usepackage{tikz}
\usepackage[T1]{fontenc}
\usepackage{etoolbox}
\usepackage{graphics}
\usepackage{siunitx}
\usepackage{hyperref}
\usepackage{float}
\usepackage{multirow}
\usepackage{mathtools}
\usepackage{bm}
\usepackage{url}

\usepackage{tikz}
\usepackage{tikz-3dplot}
\usepackage [outdir=./]{epstopdf}
\usepackage{setspace}
\usepackage{changes}

\begin{document}
\title{Electric circuit simulation of Floquet topological insulators in Fourier space}

\author{S. Sajad Dabiri}
\address{Department of Physics, Shahid Beheshti University, 1983969411 Tehran, Iran}
\author{Hosein Cheraghchi}
\email{cheraghchi@du.ac.ir}
\address{School of Physics, Damghan University, P.O. Box 36716-41167, Damghan, Iran}
\address{School of Physics, Institute for Research in Fundamental Sciences (IPM), 19395-5531, Tehran, Iran}

\date{\today}
\vspace{1cm}
\newbox\absbox
\begin{abstract}
We present a method for simulating any non-interacting and time-periodic tight-binding Hamiltonian in Fourier space using electric circuits made of inductors and capacitors. We first map the time-periodic Hamiltonian to a Floquet Hamiltonian, which converts the time dimension into a Floquet dimension. In electric circuits, this Floquet dimension is simulated as an extra spatial dimension without any time dependency in the electrical elements. The number of replicas needed in the Floquet Hamiltonian depends on the frequency and strength of the drive. We also demonstrate that we can detect the topological edge states (including the anomalous edge states in the dynamical gap) in an electric circuit by measuring the two-point impedance between the nodes. Our method paves a simple and promising way to explore and control Floquet topological phases in electric circuits.
\end{abstract}
\maketitle

\section{Introduction}
Topological insulators (TI)s are materials that have insulating bulk but support conducting edge states \cite{hasan,ando}. These edge states have a robust nature against local perturbations protected by some symmetries. These quantum materials have potential applications in various fields such as spintronics, quantum computation \cite{majorana}, etc. Albeit, materials with intrinsic topological properties are limited and it is desirable to find a route to induce topological properties in various materials. A generalization of TIs is Floquet TIs, which are created by applying a time-periodic perturbation such as a terahertz laser field to a trivial or topological insulator \cite{2011Lindner}, a semimetal \cite{oka}, or other systems. However, realizing and manipulating Floquet TIs in condensed matter systems is challenging due to some problems such as heating, which may damage the sample. Moreover, dissipation mechanisms can destroy the Floquet states \cite{sato2020} and topological features, which are unavoidable in solid-state experiments.

Recently, it has been demonstrated that topological properties can be realized in meta-materials such as electric circuits. This includes the realization of TIs \cite{tcircuits}, topological semi-metals, and higher-order TIs \cite{corner} in electric circuits. These systems are advantageous for studying topological properties because they are easy and controllable to fabricate, and topological edge states are not affected by undesired effects such as lattice imperfections and environmental interactions. The topological equivalence between $d$-dimensional static TI's with $d-1$ dimensional topological Floquet systems has been investigated by characterizing the topology of the reflection matrix from the gapless boundary states \cite{FH_static}. This result is another route to engineer Floquet systems without the need for external driving. 

In this paper, we propose a method for simulating any non-interacting and time-periodic tight-binding Hamiltonian in Fourier space using an electric circuit with time-independent inductors and capacitors. We first decompose the time-periodic Hamiltonian into a Floquet Hamiltonian (FH) using Fourier analysis and then explain how to implement it in electric circuits. The circuit will have one more dimension than the original driven system.

\section{Floquet formalism}
The Floquet theorem is a tool to study time-periodic systems. Consider a time-periodic Hamiltonian $H(t)=H(t+T)$, where $T$ is the time period, (we set $\hbar=1$). Floquet theorem states that the solution for the Schrodinger equation has the form of $ |\psi_\alpha(t)\rangle=e^{-i\varepsilon_\alpha t }|\phi_\alpha (t)\rangle $, where the Floquet quasi-modes are time-periodic $ |\phi_\alpha(t+T)\rangle=|\phi_\alpha(t)\rangle $, $\alpha$ is the band index and $\varepsilon_\alpha$ is the quasi-energy, which is defined modulo $\Omega=2\pi/T$. To transform the time-dependent Schrodinger equation into a time-independent one, we use Fourier transformation, but this increases the dimension of the Hilbert space. Using the Floquet ansatz for the eigenvalues of the Schrodinger equation, we obtain $\left[ H(t)-i\partial_{t} \right] |\phi_\alpha(t)\rangle =\varepsilon_{\alpha} |\phi_\alpha(t)\rangle$. Since Hamiltonian and Floquet quasi-modes are time-periodic, we can write the Schrodinger equation in Fourier space as \cite{kitagawa2011} the following
\begin{equation}
\mathbf{H_F} \boldsymbol{\phi}_\alpha= \varepsilon_\alpha \boldsymbol{\phi}_\alpha
\label{HF1}
\end{equation}
where
\begin{equation}
\mathbf{H_F}=
\left(\begin{array}{ccccc}
\ddots& \vdots &\vdots & \vdots & \reflectbox{$\ddots$} \\
\cdots & H^0 -\Omega&H^1 & H^2 & \cdots\\
\cdots & H^{-1} & H^0& H^1&\cdots \\
\cdots & H^{-2} &H^{-1} & H^0+\Omega & \cdots\\
\reflectbox{$\ddots$} & \vdots &\vdots & \vdots &\ddots
\end{array}\right)~
\boldsymbol{\phi}_\alpha=\begin{pmatrix}
\vdots \\
|\phi_\alpha^1\rangle \\
|\phi_\alpha^0\rangle \\
|\phi_\alpha^{-1}\rangle\\
\vdots\\
\end{pmatrix}
\label{HF}
\end{equation}
where $H^n=\frac{1}{T}\int_0^T H(t) e^{in\Omega t} dt$ and $ |\phi_\alpha^n\rangle=\frac{1}{T}\int_0^T |\phi_\alpha (t)\rangle e^{in\Omega t} dt$. The FH in Eq.~\ref{HF} is an infinite-dimensional Hermitian Hamiltonian. The eigenvalues of $\mathbf{H_F}$ form the quasi-energy band structure.
Fortunately, we can truncate the infinite-dimensional FH in Eq.~\ref{HF} for a system with a finite bandwidth to obtain an approximate solution. As noted in Refs.~\cite{handbook,prx2017}, the weight of the $m^{th}$ sideband i.e. $\langle \phi^m_\alpha|\phi^m_\alpha\rangle$ is highly suppressed for $|m|\gg \ell_m=\mathcal{W}/\Omega$. The $\mathcal{W}$ is the bandwidth, which can be taken as $\mathcal{W}=\text{max}(||H^n||)~\text{for}~n \in \mathbb{Z}$. As stated in Ref.~\cite{foa2014}, using a truncated FH with a sufficient matrix dimension, the convergence of the results is reliable giving rise to computing the time-averaged density of states, optical conductivity \cite{wu,dabiri3}, topological invariants \cite{FTIinvariants}, and other physical quantities of interest.

We can visualize the FH in Eq.~\ref{HF} as a combination of time-averaged Hamiltonians \added{($H^0$)}, with different onsite potentials \added{($\Omega$ is the difference between onsite potentials of the adjacent time-averaged replicas)}, which are coupled to each other by $H^n,~ (n\neq0)$. Here $H^n$ can be viewed as the coupling between the $n^{th}$-nearest-neighbor replicas. If $H^0$ is a tight-binding model with the dimension $d$, then the FH can be regarded as a tight-binding model with an extra dimension i.e. $d+1$. \added{Indeed, if FH in the time-independent problem shown in Eq.~\ref{HF} along the extra dimension is infinite, then Eq.~\ref{HF1} would be equivalent to the original Schrodinger equation for time-periodic Hamiltonian $H(t)$. However, truncating FH at a dimension that guarantees convergence of the results does not alter the physics of the system. It should be mentioned that for higher frequencies of the drive $\Omega$, fewer sidebands are necessary and so this method can be applied more easily.}

Let us discuss two limiting cases. First, in the high-frequency regime where $\Omega\gg\mathcal{W}$, the Floquet sidebands are effectively decoupled, known as Wannier-Stark localization \cite{okareview}, and the effect of the $n^{th}$ sideband $(n\neq0)$ can be perturbatively considered on the zero$^{th}$ sideband. In this regime, the topology of the system is described by Altland-Zirnbauer (AZ) classes \cite{ryu} for static systems in $d$ dimensions \cite{gomez}. On the other hand, if $\Omega\ll\mathcal{W}$, we are in the adiabatic regime, then we can neglect all $\Omega$'s in Eq.~\ref{HF}, and so the topology of the system is described again by the static AZ classes but in $d+1$ dimensions \cite{gomez}. In fact, beyond these two limits, topological classification in driven systems is different from static systems and shows rich features \cite{roy,FTIinvariants}.

\section{Circuit construction}
In an electric circuit, the relation between the currents and voltages is given by Kirchhoff's law \cite{tcircuits,corner,dong}.
For a time-periodic voltage of the form $V(t)=V(0) e^{i\omega t}$ with frequency $\omega$, Kirchhoff's law in Fourier space is written as
$I_p (\omega)=\sum_q J_{pq}(\omega) V_q (\omega)$
where $I_p$ ($V_p$) is the current entering from the external source (voltage) at the node $p$. In this formula, $ J (\omega)=\sum_n j_n |\chi_n\rangle\langle \chi_n|$ is the grounded Laplacian of the circuit and $j_n$, $|\chi_n\rangle$ are its eigenvalues and eigenvectors. The Laplacian is defined as \cite{dong} 
\begin{equation}
\begin{aligned}
J_{pq} (\omega)&=i\omega \left(t_{pq}(\omega)+\delta_{pq}\mu_p(\omega) \right)\\
t_{pq} (\omega)&=-C_{pq}+1/{\omega^2 L_{pq}}\\
\mu_{p} (\omega)&=C_p-1/{\omega^2 L_p}-\sum_\nu t_{p\nu}
\label{Jpq}
\end{aligned}
\end{equation}
 where $C_{p} (L_{p})$ is the capacitance (inductance) between the node $p$ and the grounded electrode, and $C_{pq} (L_{pq})$ is the capacitance (inductance) between the nodes $p$ and $q$. If we write $J_{pq} (\omega)=i\omega H_{pq}$, then $H_{pq}$ can be regarded as a tight-binding Hamiltonian, in which $t_{pq} (\omega)$ is the hopping parameter between the sites $p$ and $q$, and $\mu_p (\omega)$ denotes the onsite potential at site $p$.  It is interesting that there is a correspondence between the circuit Laplacian and a tight-binding Hamiltonian. If the electric circuit is only composed of capacitors and inductors, the $H_{pq}$ will be a Hermitian Hamiltonian, however, resistors can cause non-Hermiticity. In this simulation, the nodes play the role of atomic sites in a lattice. According to the second formula in Eq.~\ref{Jpq}, capacitors can be considered as negative hopping parameters $(-C)$ if there is no inductor while positive hoppings can be simulated by inductors $(1/\omega^2 L)$ if there is no capacitor. The complex hoppings can be implemented with the operational amplifiers \cite{calibration} or subnode \cite{dong}. The subnode method is a simulation of hopping parameters with arbitrary complex phase factors by using the elements such as capacitors, inductors, and also several subnodes defined on each node  \cite{dong}. Note that we assume that the units of the Hamiltonian $H_{pq}$ and driving frequency $\Omega$ are $n F$. It is easy to change the units of the corresponding parameters in energy.
One of the main measurable quantities is the impedance between two special nodes, which is written as \cite{tcircuits,corner}
\begin{equation}
Z_{pq}=G_{pp}+G_{qq}-G_{pq}-G_{qp}=\sum_{ n}{|\chi_{n,p}-\chi_{n,q}|^2}/{j_n}
\label{z}
\end{equation}
where $G=J^{-1}$ is the Green function of the circuit. It was shown that the two-point impedance diverges between the edges of the sample in a topological phase due to localization of the wave functions attributed to the edge states and vanishing $j_n$ \cite{tcircuits, corner}.
It is possible to implement and simulate FH as a static Hamiltonian in an electric circuit consisting of capacitors and inductors \cite{dong}. In what follows, we try to implement FH in an electric circuit for a simple model.

\begin{figure}
\includegraphics[width=\linewidth,trim={0 0 0.25cm 0}, clip]{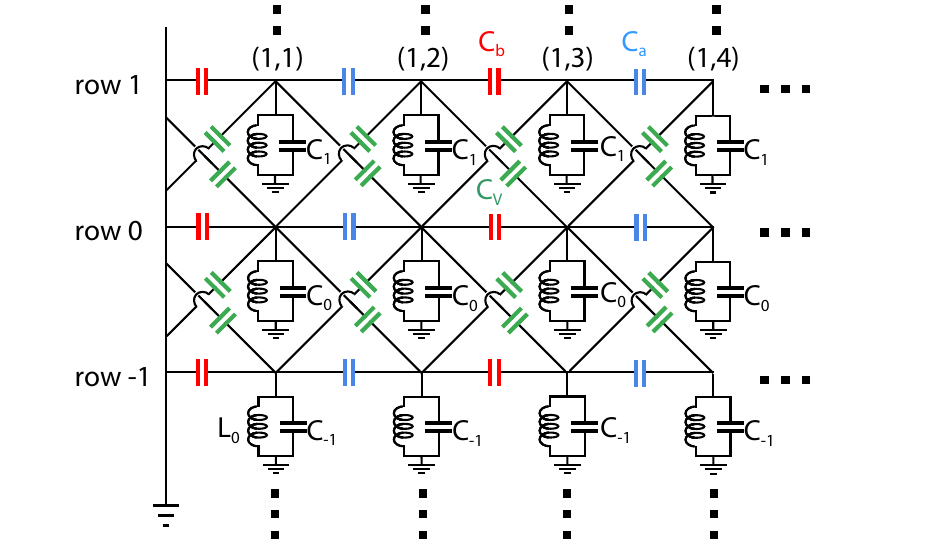} 
\caption{The driven SSH circuit. Each row shows a static SSH circuit with capacitors $C_a, C_b$. They are connected by capacitor $C_V$ which plays the role of drive. Different rows have different groundings (simulating different onsite potentials on each replica of Hamiltonian).}
\label{dssh}
\end{figure}


\section{Driven SSH circuit}
The Su-Schriffer-Heeger (SSH) model was originally proposed for the electronic states in polyacetylene \cite{sshref}. It is a one-dimensional tight-binding Hamiltonian with two atoms in each unit cell and intracell (intercell) hopping parameters as $t_a (t_b)$. In real space, it reads
$
H_{SSH}=\sum_j t_a c_{a,j}^\dagger c_{b,j}+ t_b c_{b,j-1}^\dagger c_{a,j}+h.c.
$
where $c_{a,j}^\dagger (c_{a,j})$ creates or destroys a particle on site $j$ in sublattice $a$.
For $|t_a/t_b|<1 (|t_a/t_b|>1)$, the model is in a topological (trivial) phase. Circuit simulation of this model shows that there is an impedance divergence between the ends of the chain in the topological phase \cite{tcircuits}. If we modulate the hopping parameters in a time-periodic way as $t_{a,b} (t)= t_{a,b}+2V~\text{cos} (\Omega t)$, we obtain a driven SSH model. In the momentum space, Fourier components of the corresponding time-dependent Hamiltonian can be written as  (assuming the spacing between unit cells to be unity)

\begin{equation}
H^n(k)=( t_a^{(n)}+t_b^{(n)} e^{-ik} )\sigma_++h.c.
\label{hn}
\end{equation}
where $t_{a,b}^{(0)}=t_{a,b}$, $t_{a,b}^{(\pm1)}=V$ and $t_{a,b}^{(n)}=0$ for $n\neq~0,\pm1$ and  $\sigma_+=(\sigma_x+i\sigma_y)/2$ where $\sigma_x, \sigma_y$ are Pauli matrices.

It is possible to construct an electric circuit version of FH defined in Eq.~\ref{HF} with the building blocks given by Eq.~\ref{hn} for the driven SSH model.  This circuit is displayed in Fig.~\ref{dssh}. It consists of several static SSH circuits with different groundings connected by capacitors $C_V=-V$. In this model, each replica simulated by the circuit rows contains just capacitors $C_a=-t_a$ and $C_b=-t_b$ as the hopping terms. The nodes in row $n$ are grounded by an inductor $L_0$ and capacitor $C_n$ which are set in parallel. According to Eq.~\ref{HF}, the onsite potentials on the adjacent replicas (rows in the circuit simulation) differ by $\Omega$, so that $\mu_n-\mu_{n-1}=C_n-C_{n-1}=\Omega$. Note that the grounding of the edge nodes differs from the bulk ones, however, the grounding should be such that the total elements exiting from each node in a row are identical. For example, for a node in a given row $n$, grounding is such that the onsite potential of associated Hamiltonian at this node is $\mu_n=C_a+ C_b+ 4C_V+C_n-1/( L_0 \omega^2)$.
\begin{figure}
\includegraphics[width=\linewidth,trim={0 0 0.25cm 0}, clip]{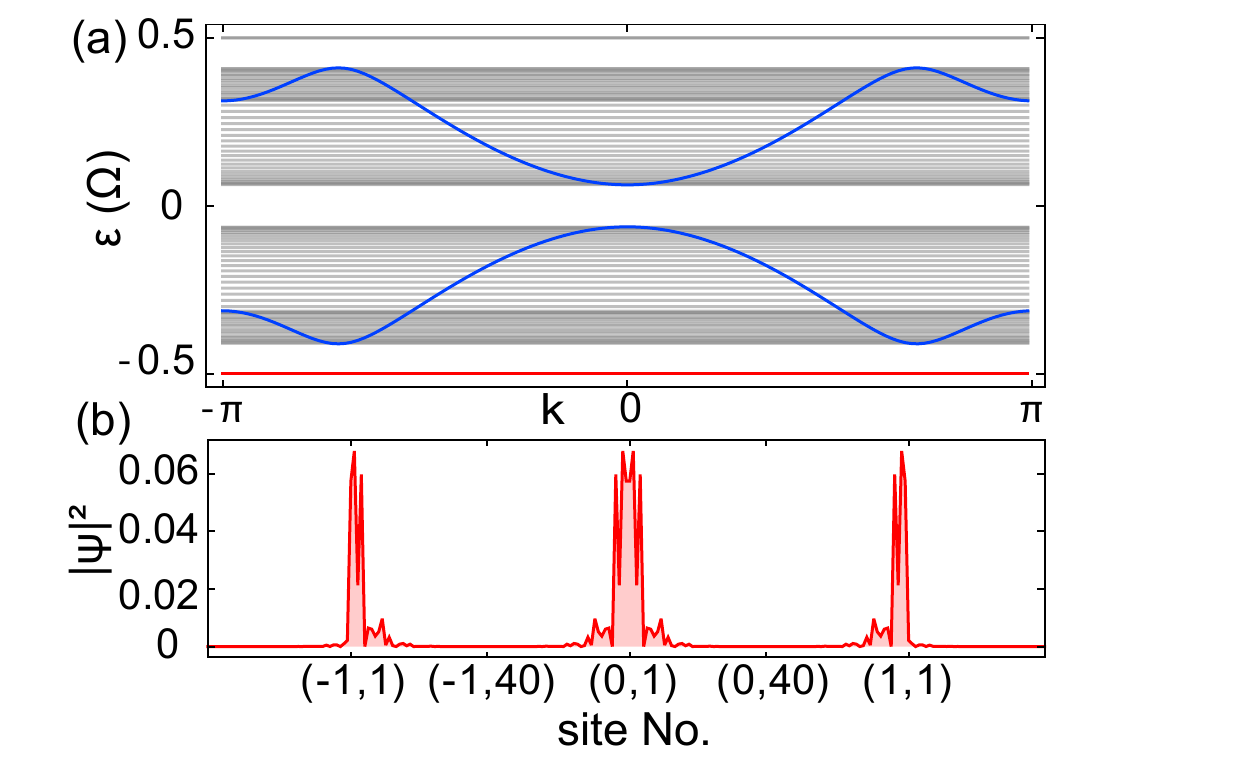} 
\caption{(a) The quasienergy dispersion of a driven SSH model with parameters $t_a=-10 ,t_b=-5 ,V=-3 ,\Omega=16 $. Blue (gray and red) lines show the bulk band structure with periodic (open) boundary conditions. (b) The square of the wave function of the midgap state at quasienery $-\Omega/2$ depicted in part (a) by red color. The site number is denoted by $(i,j)$ where the first (second) component specifies the number of row (column). }
\label{midgap}
\end{figure}

\subsection{detecting an anomalous edge mode}

The topological nature of the driven SSH model can be determined by evaluating its Zak phase \cite{foassh} which has been calculated in the supplementary material. Consider a driven SSH model with parameters  $t_a=-10, t_b=-5, V=-3,\Omega=16$. In this case, there is one resonance in its band structure between the central Floquet band ($n=0$) and the sidebands $\pm1$. Clearly, its non-driven counterpart with such parameters is in a trivial phase without any zero edge modes. However, its driven version hosts some mid-gap states at the quasi-energies $\pm\Omega/2$. The sum of Zak phases of Floquet bands below quasi-energy 0 ($\Omega/2$) equals 0 ($\pi$) modulo 2$\pi$. Fig.~\ref{midgap}(a) shows the band structure of the driven SSH model. The blue (gray or red) lines correspond to states with periodic (open) boundary conditions. The red line indicates the mid-gap states (at $-\Omega/2$) whose squared wave function is plotted in Fig.~\ref{midgap}(b). It is seen that the wave function is mainly localized at the edges of the side-bands $-1$ and $0$ because this mid-gap state results from the resonance between these two sidebands. The wave function decays exponentially to the bulk with an exponent proportional to the associated gap \cite{foassh}. However, by increasing the drive amplitude, the wave function will have non-negligible values on the edges of other side-bands too, however, this effect is highly suppressed for the sidebands which are farther from the sidebands $-1$ and $0$.
\begin{figure}
\includegraphics[width=\linewidth,trim={0 0 0.25cm 0}, clip]{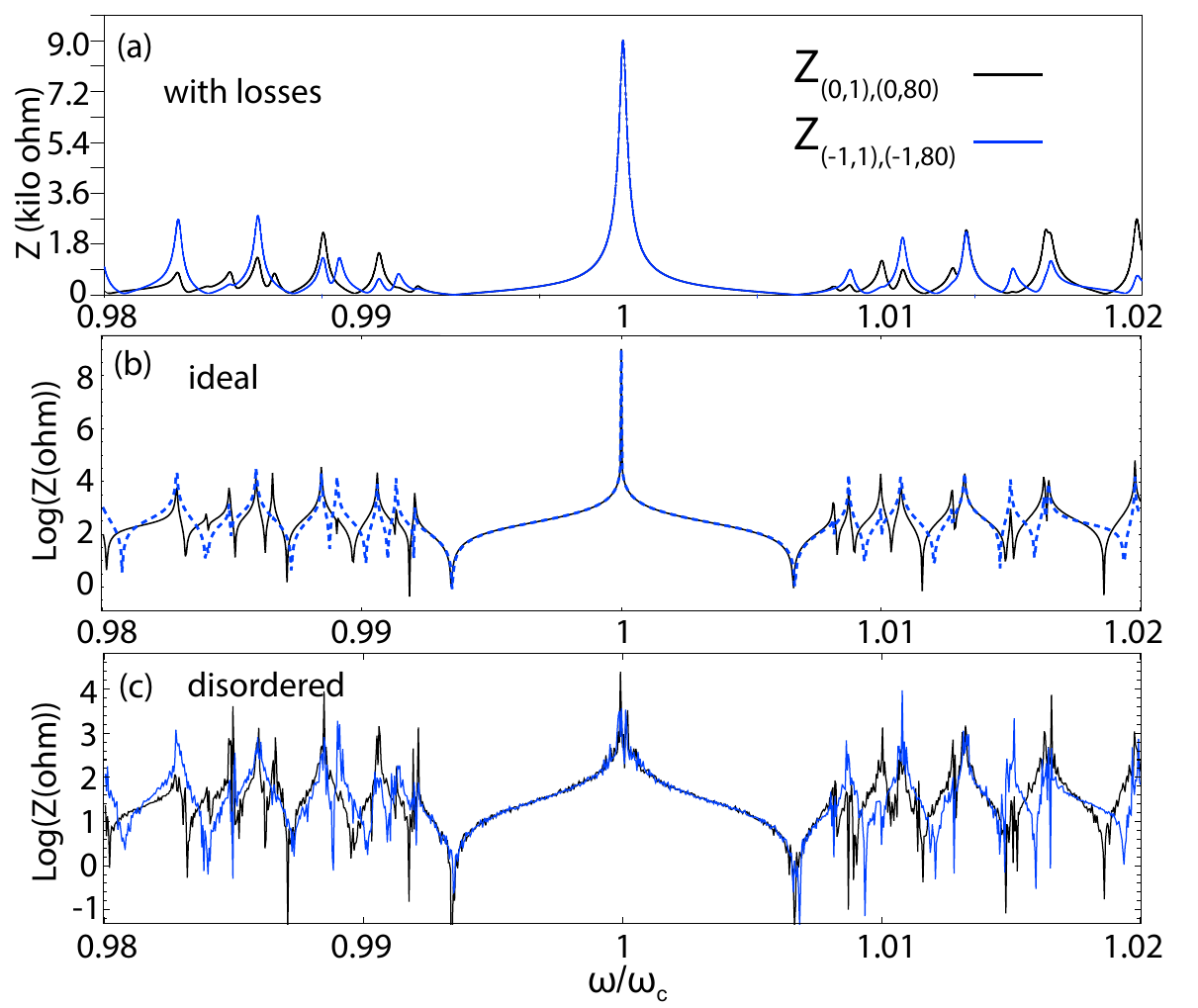} 
\caption{The impedance between the ends of the 0$^{th}$ (-1$^{th}$) row, which is specified by black (blue) color. The circuit parameters are $C_a=10~\text{nF} ,C_b=5~\text{nF}, C_V=3~\text{nF}, C_{-4}=\Omega/2, C_n-C_{n-1}=\Omega, \Omega=16~\text{nF}, L_0=10~\mu \text{H}, R_{dc}=2.8~\text{mOhm}$. The result of simulation with and without $R_{dc}$ is shown in parts (a) and (b) respectively. \added{(c) The impedance in the presence of $2\%$ disorder in all elements including $C_a, C_b, C_V, L_0, \Omega$.} }
\label{imp}
\end{figure}

As indicated by the red line in Fig.~\ref{midgap}, a mid-gap state is pinned to the quasi-energy $-\Omega/2$. To get a divergence in the impedance, we should adjust the chemical potential to $-\Omega/2$. Equivalently, we tune the frequency such that the onsite potentials at different rows become $\mu_0(\omega_c)=\Omega/2, \mu_n-\mu_{n-1}=\Omega$ (see the third formula of Eq.~\ref{Jpq}). Let us consider $(2\times4+1)=9$ sidebands in our numerical calculations and set $C_{-4}=\Omega/2$, $C_n-C_{n-1}=\Omega$. Then, it is expected that at the frequency of $\omega_c$, there is a divergence for the impedance between the ends of the chain in $-1$ and $0$ rows. The resonant frequency $\omega_c$ is obtained using Eq.~\ref{Jpq} which yields $\mu_0(\omega_c)=C_a+C_b+4C_V+C_0-\frac{1}{L_0 \omega_c^2}=\frac{\Omega}{2}$. We assume 80 sites in each row. Fig.~\ref{imp} represents the two-point impedance $Z_{(0,1),(0,80)}$ between the nodes $(0,1)$ and $(0,80)$. Furthermore, to compare impedance divergence in different rows, $Z_{(-1,1),(-1,80)}$ is also depicted by the blue line in this figure. The impedance with and without serial resistances attached to the inductors are represented in Figs.~\ref{imp}(a),(b), respectively. To verify our results, we also checked them by LTspice software for the impedance calculation in the presence of losses. The divergence in impedance occurs at the frequency $\omega_c$ which is clearly seen in Fig.~\ref{imp}, albeit the peaks are broadened in the presence of losses. It is also checked that this divergence also occurs for $Z_{(-1,1),(0,80)}$ at $\omega_c$ but not for $Z_{(-1,1),(0,1)}$ or $Z_{(-1,80),(0,80)}$. \added{To demonstrate the protection of this peak against disorder, the impedance is calculated in the presence of $2\%$ disorder in  Fig.~\ref{imp}(c). The impedance peak is robust after introducing such weak disorder $\lesssim 2\%$ to the system, however, the position of the impedance peak may be shifted slightly and its height is also altered. See supplementary material for the results for different disorder strengths.}

\begin{figure*}
\includegraphics[width=\linewidth,trim={0 0 0.25cm 0}, clip]{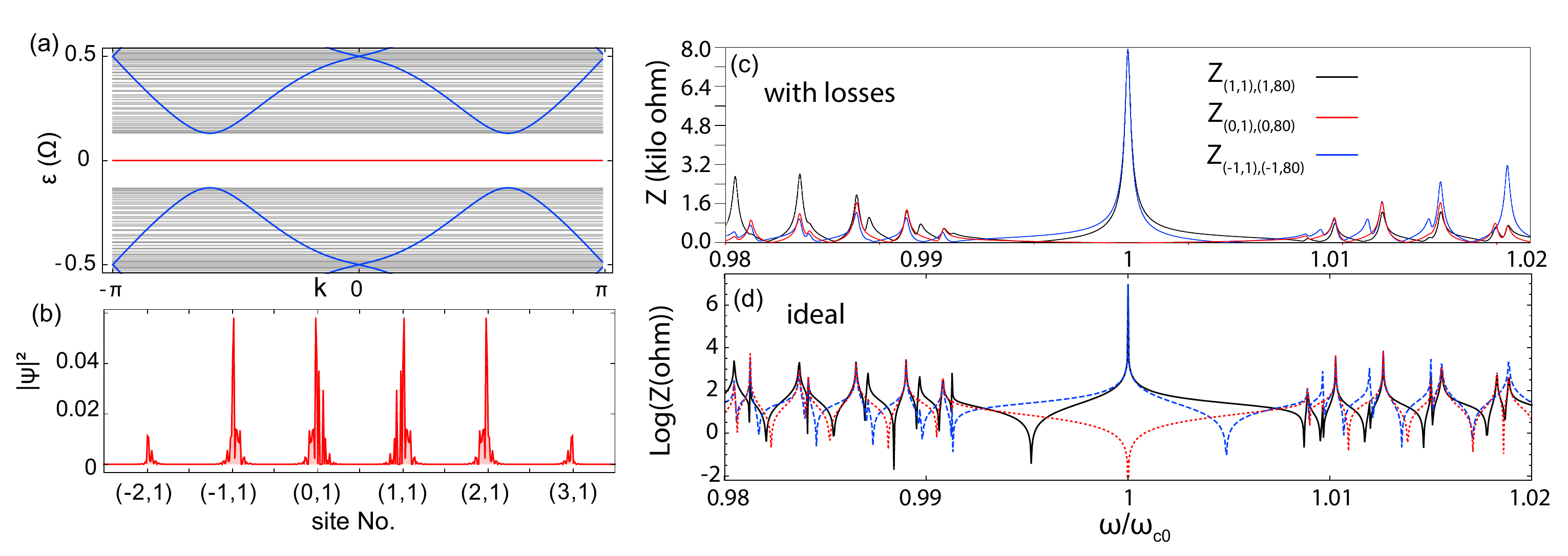} 
\caption{(a) bands dispersion for tight binding FH with parameters $|t_a|=C_a=10~\text{nF}, |t_b|=C_b=5~\text{nF}, |V|=C_V=4~\text{nF}, \Omega=10~\text{nF}$. Blue bands show the bulk band structure with periodic boundary conditions and gray and red bands show the dispersion with open boundary conditions (80 sites in each row are used). (b) the square of the wave function for the midgap states at zero quasienergy which is denoted by red color in (a). (c),(d) results of two-point impedance between ends of three rows with and without serial resistance $R_{dc}$ respectively. The parameters are the same as (a) and $L_0=10~\mu H, R_{dc}=2.8~\text{mOhm}$. }
\label{om1}
\end{figure*}

\subsection{Detecting midgap states at zero quasienergy}
Now we present the results for the emergence of midgap states at zero quasienergy. Fig.~\ref{om1} (a) shows the band dispersion of the driven SSH model obtained from Eqs.~\ref{HF} and \ref{hn} for parameters $t_a=10, t_b=5, V=4, \Omega=10$. As can be seen, there are midgap states at zero quasienergy which are depicted by the red color. These midgap states are mainly due to the resonance between the $-1^{th}$ and $1^{th}$ sidebands. It is evident that these midgap states are absent for non-driven case i.e. $V=0$, because $t=t_a/t_b>1$. Fig.~\ref{om1} (b) shows the square of the wave function for midgap states. It is seen that the amplitude of the wave function has the highest values at the ends of $-1^{th}$ and $1^{th}$ side bands. Fig.~\ref{om1}(c),(d) shows the two-point impedance between the ends of $-1^{th}$, $0^{th}$, $1^{th}$row with black, red, and blue color, respectively. The parameters are $C_a=10~\text{nF}, C_b=5~\text{nF}, C_V=4~\text{nF}, \Omega=10~\text{nF}, L_0=10~\mu H$ and the resonance frequency $\omega_c$ is obtained from the equation $\mu_0=C_a+C_b+4C_V+C_0-\frac{1}{L_0\omega^2_{c0}}=0$ where $C_0=4.5 \Omega$. In Fig.~\ref{om1}(c) a serial resistance for inductors $R_{dc}=2.8~\text{mOhm}$ has been assumed which is usually present in experimental setups. As it is clear from Figs.\ref{om1}(c),(d) there is a peak for impedance between ends of $-1^{th}$ and $1^{th}$ row, but not for $0^{th}$ row.

Although we discussed the two-point impedance, one-point impedance is another measurement. It is defined as $Z_{p}=G_{pp}=\sum_{ n}\frac{|\chi_{n,p}|^2}{j_n}$. In order to measure this quantity, one should insert a current from the ground by an external current source to a node and measure the voltage difference between that node and the ground. The ratio of voltage and current is the one-point impedance. As it is clear from its definition, it has a behavior similar to the amplitude of the wave function which has the smallest eigenvalue. So we expect that at the resonance frequency, one gets a one-point impedance pattern similar to the wave function of midgap states. For example, for the parameters used in Fig.~\ref{om1} one-point impedance has its highest value at edges of $-1^{th}$ and $1^{th}$ rows and decays exponentially for the bulk nodes.

\subsection{high and low-frequency limits}
Now let us focus on high and low-frequency regimes. In high-frequency regime $\Omega\gg\mathcal{W}$, the effect of other sidebands projected on the zero$^{th}$ sideband can be perturbatively considered. Then essentially an effective  Hamiltonian is proposed which captures the topological properties of the system. If $U(T)$ denotes the evolution operator of the system in stroboscopic times, then the effective Hamiltonian is defined as $U(T)=\mathcal{T}\int_0^T e^{-iH(t)}dt=e^{-iH_{\text{eff}}T}$. There are some expansions of this effective Hamiltonian in inverse powers of frequencies known as van-Vleck and Brillouin-Wigner expansions \cite{bw}. For example, the effect of high-frequency light on topological insulator thin films is analyzed in Refs.\cite{dabiri1,dabiri2}. If the effective Hamiltonian $H_{\text{eff}}$ is topologic with a non-trivial winding number, then the model displayed in Fig.~\ref{dssh} would possess zero modes localized mostly at the ends of the zero$^{th}$ row.

\begin{figure}
\includegraphics[width=\linewidth,trim={0 0 0.25cm 0}, clip]{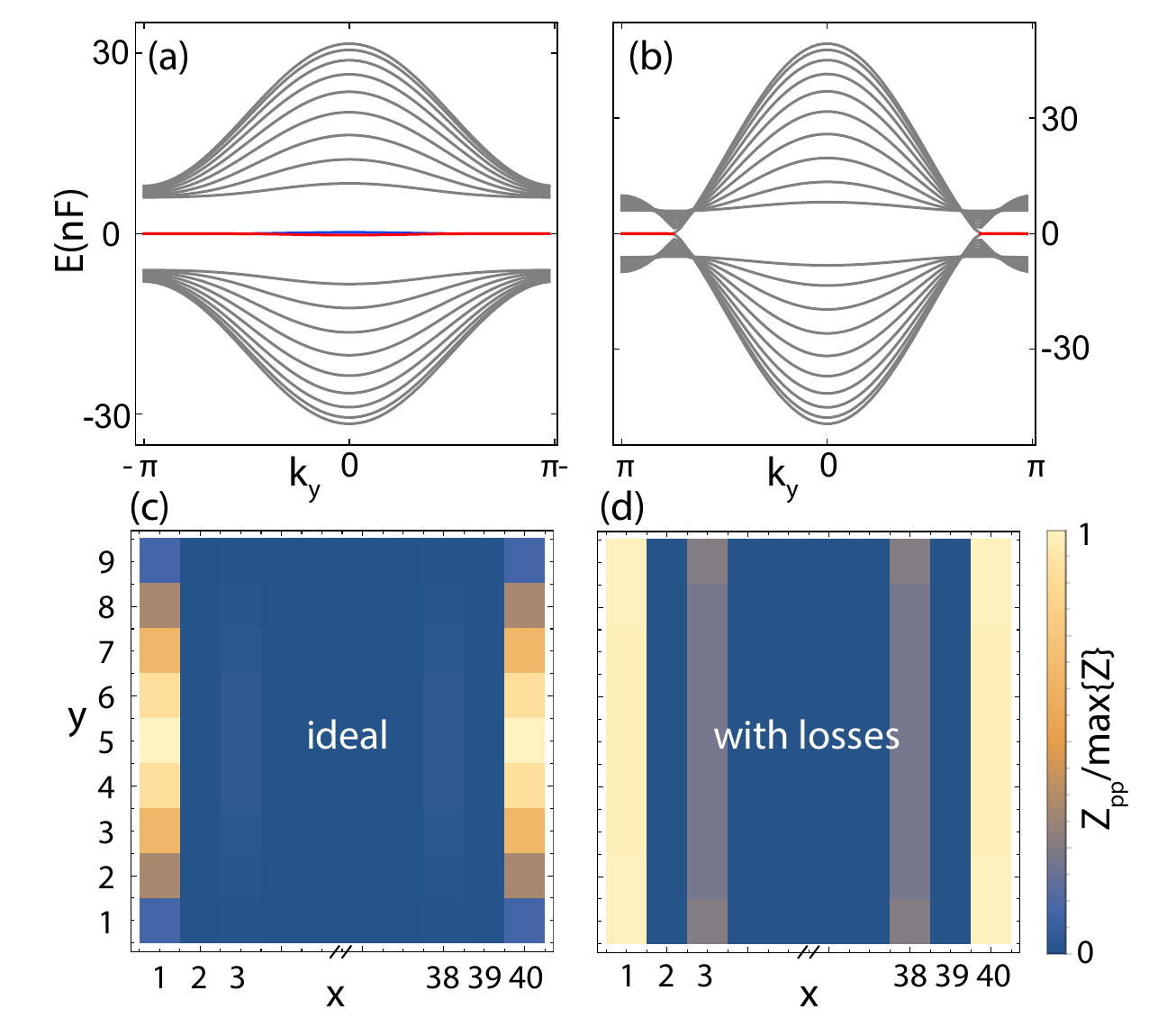} 
\caption{Dispersion relation of nanoribbon version of Hamiltonian in Eq.~\ref{hadia} for the parameters (a) $|t_a|=C_a=7~\text{nF}, |t_b|=C_b=13~\text{nF}, 2 |V|=2C_V=6~\text{nF}$ and (b) $t_a=13~\text{nF}, t_b=7~\text{nF}, 2V=15~\text{nF}$ with open (periodic) boundary condition along $x(y)$. \added{(c) The distribution of one-point impedances at resonant frequency $\omega_{c1}$ for driven SSH circuit when we set $C_n=0$ in Fig.~\ref{dssh} with parameters as used for the panel (a) and  $L_0=10~\mu \text{H}$ with open boundary conditions along $x$ and $y$. (d) the same as (c) with extra losses $R_{dc}=2.8~\text{mOhm}$.} }
\label{adia}
\end{figure}


On the other hand, in low-frequency regime $\Omega\ll\mathcal{W}$, one can set the groundings in Fig.~\ref{dssh} identically for all nodes, i.e. $C_n=0$. Assuming $\Omega t\rightarrow k_y,k\rightarrow k_x$ in the driven SSH model, then the Hamiltonian in this adiabatic regime can be written as 

\begin{equation}
\begin{aligned}
H_{\text{adia}}(k_x,k_y)=&[t_a +2 V \cos (k_y) \\
&+(t_b +2V \cos (k_y)) e^{-ik}] \sigma_++h.c. 
\label{hadia}
\end{aligned}
\end{equation}
This Hamiltonian resembles the Hamiltonian of the zigzag graphene model introduced in Ref.~\cite{tcircuits}. Although the Hamiltonian \ref{hadia} has zero Chern number due to the presence of time-reversal symmetry, nearly flat edge bands are formed in its band structure. These edge states can be detected by the impedance measurement \cite{tcircuits}.

The nearly flat edge bands are present (absent) if $|\frac{t_a+2V \cos{k_y}}{t_b+2V \cos{k_y}}|<1 (|\frac{t_a+2V \cos{k_y}}{t_b+2V \cos{k_y}}|>1)$ which originates from the winding number at fixed $k_y$ as $k_x$ varies through $(-\pi,\pi)$. See Ref.~\cite{flatezawa} for topological arguments about these edge states with nearly flat bands. The band structure of a nanoribbon version of the Hamiltonian~\ref{hadia} for the given parameters $t_a=7,t_b=13, 2V=6 (t_a=13,t_b=7, 2V=15)$ is presented in Fig.~\ref{adia} a (b), respectively. The edge states localized on the left and right edges are shown by the red and blue lines. Clearly, one case is an insulator with the edge states appearing at all momenta and the other case is a semi-metal with edge states appearing at some special range of momenta $|k_y|>\arccos(-2/3)$.

\added{We also calculate the distribution of one-point impedances $Z_{pp}=G_{pp}=\sum_{ n}\frac{|\chi_{n,p}|^2}{j_n}$ in a lattice with the size of $40\times9$ and for parameters that was used in Fig.~\ref{adia} (a) at the resonant frequency $\omega_{c1}=1/\sqrt{(C_a+C_b+4 C_V)L_0}$ with and without losses in Fig.~\ref{adia} (c),(d), respectively. In Fig.~\ref{adia} (d)  serial resistances $R_{dc}$ are assumed for the inductors.  The impedance peaks at two edges of the sample in Fig.~\ref{adia} (c),(d)  correspond to the edge states with nearly flat bands, giving rise to a weak topological phase.}

\section{Conclusion}
In summary, we have shown that every non-interacting and tight-binding Floquet Hamiltonian (FH) can be simulated in an electric circuit. \added{Although we showed this correspondence for a 1D model, thanks to the easiness of the simulation of couplings in electrical circuits, the driven models in higher dimensions (even greater than 3) can be simulated as well. For a $d+1$-dimensional FH, each replica in Eq.~\ref{HF} is a $d$-dimensional model ($H^0$ with different onsite potentials) instead of a 1D chain, and replicas are coupled by $H^n~(n\neq0)$}. There are some aspects that distinguish the FH from all conventional static topological systems, like TIs, higher order TIs, topological semimetals, etc. First, the topological invariants which describe the topology of FH are different from those that are used for conventional static topological systems \cite{roy,FTIinvariants}. Second, the localization of the edge states is different. For example, resonances between the central Floquet bands and the other sidebands induce the edge states with a localization that depends on the amplitude of the drive. Third, unlike conventional static systems which mainly have one gap located at zero energy, the FH has several gaps in which topological edge states emerge. Our work paves the way for studying and detecting the Floquet topological phases in experimental setups.

\section{supplementary material}
Supplementary material includes the details about the numerical calculation of the Zak phase, the effect of different amounts of disorder on two-point impedances, a clarification about the difference between the groundings of edge, corner, and bulk nodes in a driven SSH circuit, and a circuit construction of an SSH model when sublattice potentials are modulated in time.

\section{Acknowledgements}
We highly appreciate Alina Rozenblit and Meysam Zareiee for the discussion on circuit simulation.

\section{data availability}
The data that support the findings of this study are available within the article and its
supplementary material.

\pagebreak

\onecolumngrid
\begin{center}
\textbf{\large Supplementary material for "Electric circuit simulation of Floquet topological insulators in Fourier space"}\\[.2cm]
S. Sajad Dabiri,$^{1}$ Hosein Cheraghchi,$^{2,3}$ \\[.1cm]
{\itshape ${}^1$Department of Physics, Shahid Beheshti University, 1983969411 Tehran, Iran\\
${}^2$School of Physics, Damghan University, P.O. Box 36716-41167, Damghan, Iran\\
${}^3$School of Physics, Institute for Research in Fundamental Sciences (IPM), 19395-5531, Tehran, Iran\\}
(Dated: \today)\\[1cm]
\end{center}

\setcounter{equation}{0}
\setcounter{figure}{0}
\setcounter{table}{0}
\setcounter{page}{1}
\renewcommand{\theequation}{S\arabic{equation}}
\renewcommand{\thefigure}{S\arabic{figure}}
\renewcommand{\bibnumfmt}[1]{[S#1]}
\renewcommand{\citenumfont}[1]{S#1}

\section{numerical calculation of Zak phase of driven SSH model }
\begin{figure}
\includegraphics[width=\linewidth,trim={0 0 0.25cm 0}, clip]{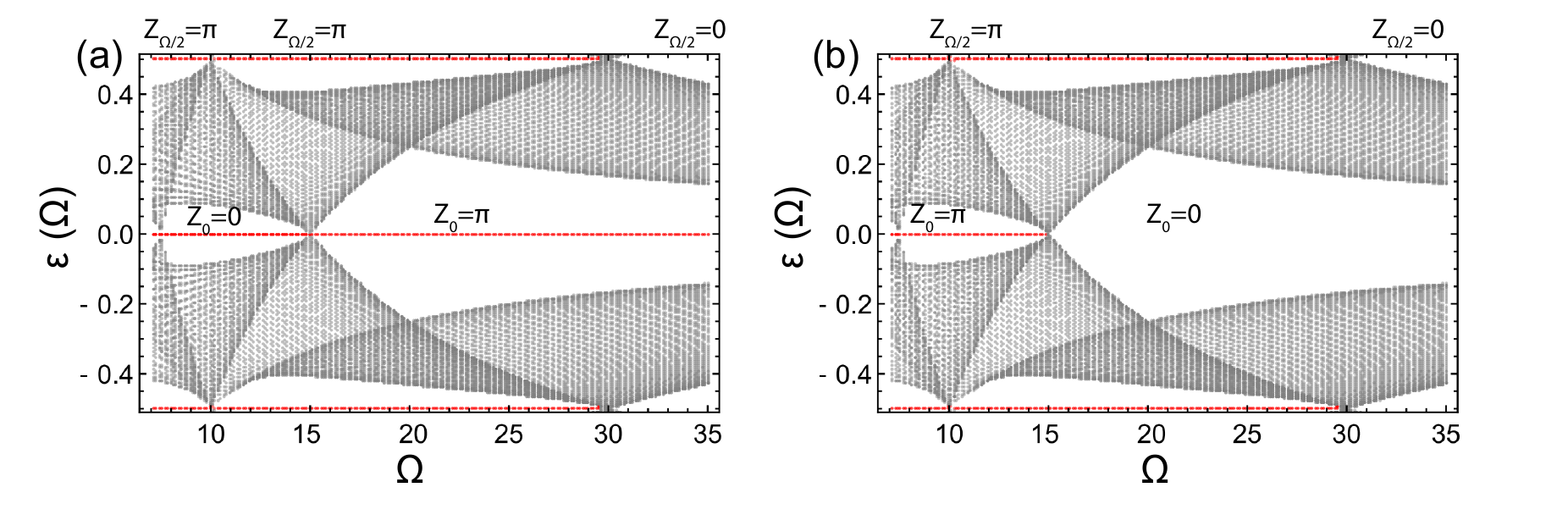}
\caption{ Bands dispersion for the tight binding Floquet Hamiltonian with parameters (a) $t_a=5, t_b=10, V=3$ and (b) $t_a=10, t_b=5, V=3$ as a function of the drive frequency with an open boundary condition. The edge modes are shown in the red color. The Zak phase of each gap $Z_0, Z_{\Omega/2}$ is denoted in this figure. }
\label{zak}
\end{figure}

We can calculate the Zak phase for each gap, i.e. $Z_0, Z_{\Omega/2}$ for the truncated FH, in principle analytically but in practice numerically. Because for a large number of bands, the analytical calculation of integrals is very difficult. In Fig.~\ref{zak}(a),(b) we show the numerical result of band dispersion of the truncated FH for the driven SSH model with parameters  (a) $t_a=5, t_b=10, V=3$ and (b) $t_a=10, t_b=5, V=3$ as a function of the drive frequency with an open boundary condition. The edge modes are denoted by the red color and the zak phase of each gap is calculated. When each gap closes, the Zak phase of that gap can change. An interesting point emerges in this figure. For frequencies lower than $\Omega=15$ in Fig.~\ref{zak}(a), the Zak phase of the gap $\varepsilon=0$ is trivial but there are four edge modes in this gap.

The numerical method for calculating the Zak phase is as follows. The Brillouin zone must be discretized to $n$ pieces. Then a phase difference between the wavefunctions of a specific band of FH on adjacent sites in k-space should be calculated
\begin{equation}
\begin{aligned}
U_l=\frac{\langle \psi_l | \langle \psi_{l+1}\rangle}{|\langle \psi_l | \langle \psi_{l+1}\rangle|}
\end{aligned}
\end{equation}
Where $\psi_l$ is the eigenfunction of FH related to a specific band at $k_l$. The numerical Zak phase for that band is obtained as
\begin{equation}
\begin{aligned}
\Phi_Z=\sum_{l=1}^{n} U_l~\text{mod}~2\pi
\end{aligned}
\end{equation}
So the Zak phase for a given gap is equal to the summation of all Zak phases of bands below that gap modulo $2\pi$.

\begin{figure}
\includegraphics[width=\linewidth,trim={0 0 0.25cm 0}, clip]{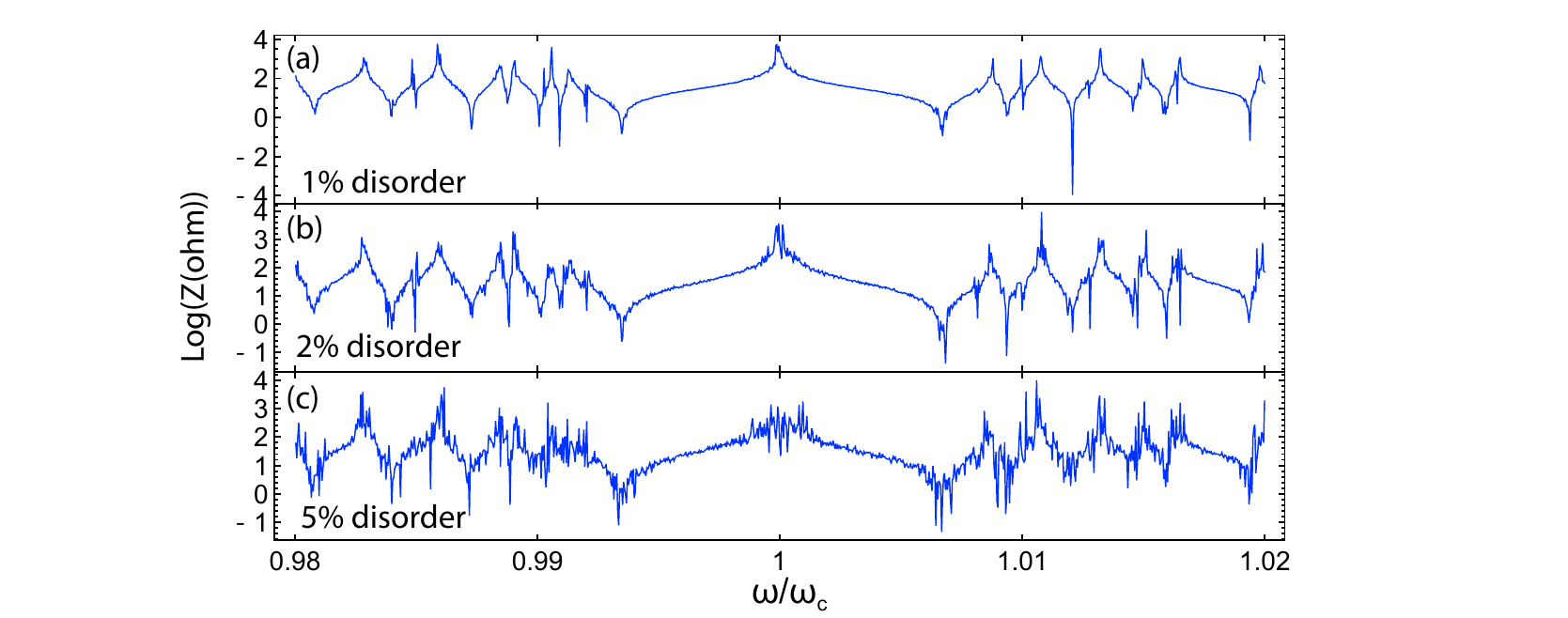}
\caption{The impedance between the ends of the -1$^{th}$ row for driven SSH circcuit in topological phase. The circuit parameters are $C_a=10~\text{nF}-\delta \%, C_b=5~\text{nF}-\delta \%, C_V=3~\text{nF}-\delta \%, \Omega=16~\text{nF}-\delta \%, L_0=10~\mu H-\delta \%,  C_{-4}=\Omega/2, C_n-C{n-1}=\Omega$ with $\delta=1, 2, 5$ in panels (a), (b), (c), respectively. }
\label{disfig}
\end{figure}

\section{effect of disorder on impedances }
In this section, we investigate the effect of different amounts of disorder on impedances in the topological phase. Consider the driven SSH model in the phase with anomalous midgap states at $\Omega/2$ as in Fig.3 of the main text. Let us assume $\delta \%$ of disorder and set the parameters as $C_a=10~\text{nF}-\delta \%, C_b=5~\text{nF}-\delta \%, C_V=3~\text{nF}-\delta \%, \Omega=16~\text{nF}-\delta \%, L_0=10~\mu H-\delta \%,  C_{-4}=\Omega/2, C_n-C{n-1}=\Omega$. The impedance between ends of the $-1^{th}$ row of a circuit with size $80\times 9$ i.e. $Z_{(-1,1),(-1,80)}$ for different values of disorder are shown in Fig.~\ref{disfig}. Although the frequency and height of the peaks are affected by the disorder, the peaks are persistent against a weak amount of disorder $\delta \lesssim 2\%$. So a weak disorder can not change the results significantly and the models introduced in this paper with the topological features seem applicable in real experimental setups.

\begin{figure}
\includegraphics[width=\linewidth,trim={0 0 0.25cm 0}, clip]{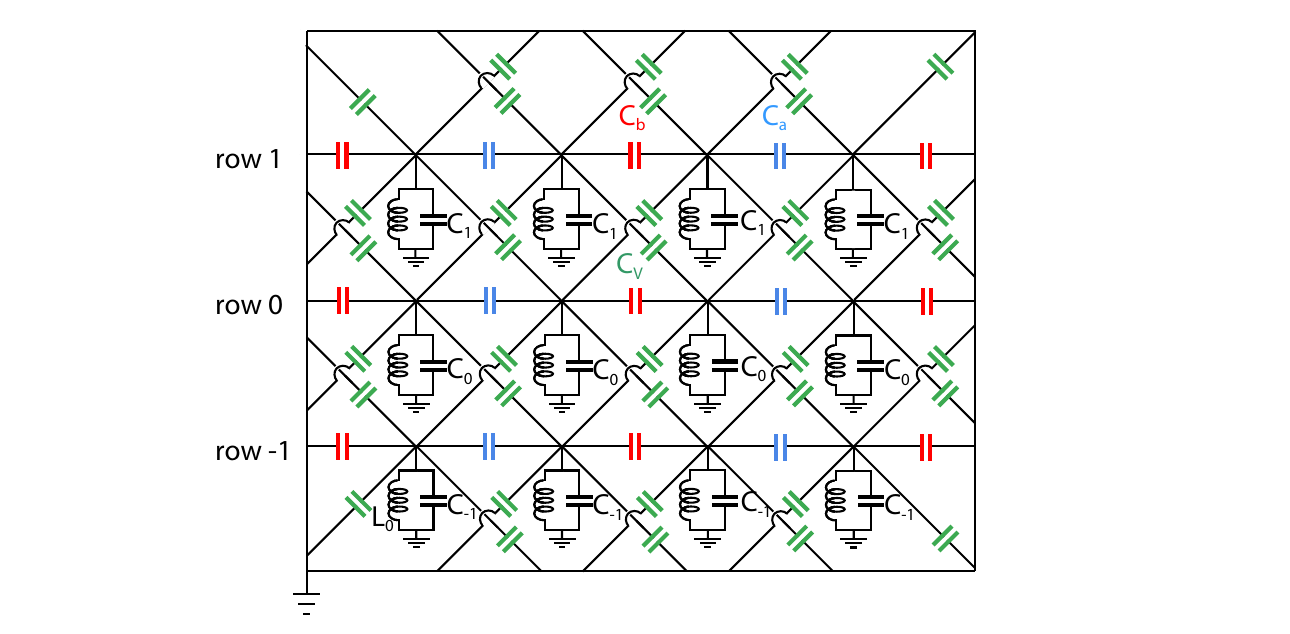}
\caption{Schematic representation of driven SSH circuit of size $3\times4$.}
\label{supcirc}
\end{figure}
\section{grounding of edge nodes }
As stated in the main text, the grounding of the bulk nodes are different from edge and corner nodes. In Fig.~\ref{supcirc} we show a finite circuit of size $3\times4$ explicitly. Note that the groundings are assumed such that the total elements exiting from each node in a row are identical. So, more elements should be connected to the ground at corner and edge nodes in order to compensate for the lack of neighbor nodes. The onsite potential for each node at row $n$ is equal to  $\mu_n=C_a+ C_b+ 4C_V+C_n-1/( L_0 \omega^2)$ which is compatible with Fig.~\ref{supcirc}.

\begin{figure}
\includegraphics[width=0.6\linewidth,trim={0 0 0.25cm 0}, clip]{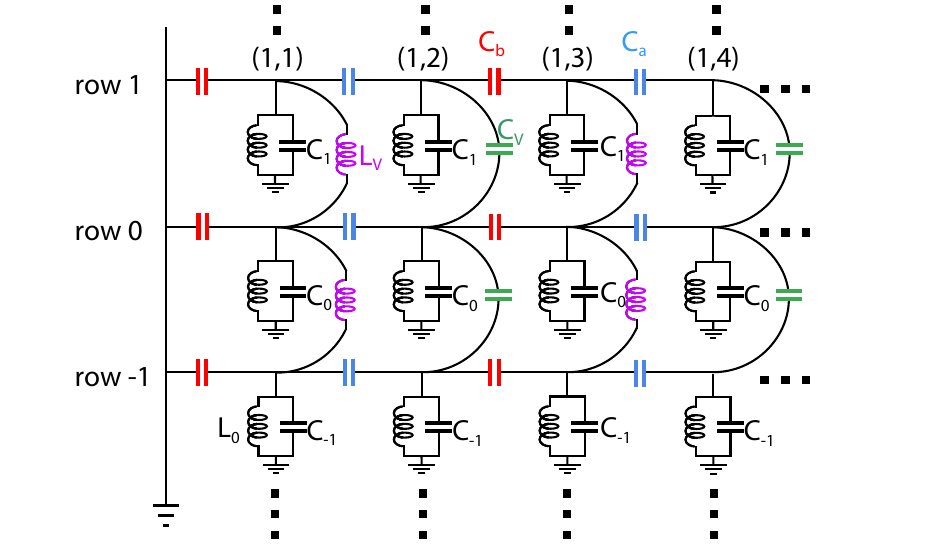}
\caption{Schematic representation of the circuit for SSH model with modulated onsite potentials whose Hamiltonian is represented in \ref{drivenssh2}}
\label{ssh2}
\end{figure}

\section{modulating the onsite potential in SSH model }
Another possibility to obtain a periodically driven SSH model is to add a time-periodic staggered sublattice potential. The time-periodic Hamiltonian in $k$-space would be 
\begin{equation}
\mathcal{H}(k,t)=[(t_a 
+t_b e^{-ik})\sigma_++h.c. ]+2V\cos(\Omega t)\sigma_z
\label{drivenssh2}
\end{equation}
The Fourier components of the above Hamiltonian are as follows
$$\mathcal{H}^0=[(t_a 
+t_b e^{-ik})\sigma_++h.c. ]~~\mathcal{H}^{\pm1}=V \sigma_z$$
The resulting circuit is shown in Fig.~\ref{ssh2} which is composed of several static SSH circuits which are connected by capacitors and inductors. The node $(i,2n)$ will be connected to the node $(i+1,2n)$ by a capacitor $C_V$ and the node  $(i,2n+1)$ will be connected to the node $(i+1,2n+1)$ by an inductor $L_V$, where at resonant frequency we should have $\frac{1}{L_V\omega_{c3}^2}=C_V$. It should be mentioned that since in Fig.~\ref{ssh2} we used inductors to simulate positive hopping parameters, the magnitude of this hopping will depend on the frequency through $\frac{1}{L_V\omega^2}$. So, if we want to detect a resonance in the spectrum of Hamiltonian, we should tune frequency such that simultaneously the chemical potential is equal to the energy of edge modes and the positive hopping parameter is equal to $C_V$. If this tuning is not possible, we can use a parallel capacitor $C'_V$ to all inductors $L_V$ such that at resonant frequency $-C'_V+\frac{1}{L_V\omega^2}=C_V$.

\end{document}